\documentclass[prl,amsmath,amssymb,reprint,superscriptaddress,nofootinbib,
tightenlines,nobibnotes,longbibliography]{revtex4-2}

\usepackage{graphicx,txfonts}
\usepackage{color}
\usepackage{booktabs}
\usepackage{array}
\usepackage{upgreek}
\usepackage[colorlinks=true, citecolor=red, linkcolor=blue, urlcolor=blue ]{hyperref}

\newcommand{\sst}[1]{\scriptscriptstyle{#1}}

\newcommand{\expo}{\mbox{e}}
\newcommand{\vecF}{\mbox{\boldmath$F$}}
\newcommand{\vecz}{\mbox{\boldmath$\hat{z}$}}

\newcommand{\veps}{\varepsilon}

\newcommand{\vecy}{\mbox{\boldmath$\hat{y}$}}

\newcommand{\vecdel}{\mbox{\boldmath$\delta$}}

\newcommand{\vecU}{\mbox{\boldmath$U$}}

\newcommand{\vecV}{\mbox{\boldmath$V$}}

\newcommand{\dvecV}{\dot{\mbox{\boldmath$V$}}}

\newcommand{\vecd}{\mbox{\boldmath$d$}}

\newcommand{\vecL}{\mbox{\boldmath$L$}}

\newcommand{\icomps}{\mbox{\scriptsize i}}
\newcommand{\icomp}{\mbox{i}}

\newcommand{\dth}{\mbox{\boldmath$\dot{\theta}$}}
\newcommand{\ddth}{\mbox{\boldmath$\ddot{\theta}$}}

\newcommand{\Lpar}{\Lambda_{\sst \Vert}}
\newcommand{\Lperp}{\Lambda_{\sst \perp}}
\newcommand{\Lr}{\Lambda_{r}}

\newcommand{\Vprop}{V_{\mbox{\scriptsize prop}}}

\newcommand{\PreserveBackslash}[1]{\let\temp=\\#1\let\\=\temp}
\newcolumntype{C}[1]{>{\PreserveBackslash\centering}p{#1}}
\newcolumntype{R}[1]{>{\PreserveBackslash\raggedleft}p{#1}}
\newcolumntype{L}[1]{>{\PreserveBackslash\raggedright}p{#1}}

\begin{document}

\title{Non inertial acoustic propulsion of bimetallic rods}

\title{Purely viscous acoustic propulsion of bimetallic rods}

\author{Jeffrey McNeill}
\affiliation{Department of Chemistry, University of Pennsylvania, Philadelphia, PA 19104-6323, USA}
\author{Nathan Sinai}
\affiliation{Department of Chemistry, University of Pennsylvania, Philadelphia, PA 19104-6323, USA}
\author{Justin Wang}
\affiliation{Department of Chemistry, University of Pennsylvania, Philadelphia, PA 19104-6323, USA}
\author{Vincent Oliver}
\affiliation{Department of Chemistry, University of Pennsylvania, Philadelphia, PA 19104-6323, USA}
\author{Eric Lauga}
\affiliation{Department of Applied Mathematics and Theoretical Physics, Centre for Mathematical
Sciences, University of Cambridge, Wilberforce Road, Cambridge CB3 0WA, UK}
\author{Fran\c cois Nadal}
\email{f.r.nadal@lboro.ac.uk} 
\affiliation{Wolfson  School  of  Mechanical,  Electrical  and  Manufacturing  Engineering, 
Loughborough  University, Loughborough,  LE11  3TU,  United  Kingdom}
\author{Thomas E. Mallouk} 
\affiliation{Department of Chemistry, University of Pennsylvania, Philadelphia, PA 19104-6323, USA}
\affiliation{International Centre for Materials Nanoarchitectonics (WPI-MANA), National Institute
for Materials Science (NIMS), 1-1 Namiki, Tsukuba, Ibaraki 305-0044, Japan}
\date{\today}

\begin{abstract}
Synthetic microswimmers offer models for cell motility and their tunability makes them promising candidates for biomedical applications. Here we measure the acoustic propulsion of bimetallic micro-rods that, when trapped at the nodal plane of a MHz acoustic resonator, swim with speeds of up to 300\,$\upmu$m\,s$^{-1}$. While past acoustic streaming models predict speeds that are more than one order of magnitude smaller than our measurements, we demonstrate that the acoustic locomotion of the rods is driven by a viscous, non-reciprocal mechanism relying on shape anisotropy akin to that used by swimming cells and that reproduces our data with no adjustable parameters.
\end{abstract}

\maketitle


Biological and synthetic nano- and micro-swimmers operate at low Reynolds number,  and their study has been a place to  discover and leverage new physics~\cite{Purcell1977,Lauga2009}. Since their introduction in 2004~\cite{Paxton2004}, synthetic swimmers have been extensively studied, 
both to understand the  fundamental physics of active fluids~\cite{marchetti_review} and also for potential applications in biomedicine~\cite{Wu2018}, materials 
science \cite{Lavergne2019,Lozano2019}, environmental remediation~\cite{Ying2019}, 
and analytical science~\cite{Wang2018}. These swimmers 
range in size from tens of nanometers to micrometers, and can be powered by catalysis~\cite{Ma2015}, 
magnetic~\cite{Gao2010,Xie2019,Dreyfus2005} and electric fields \cite{Yan2016,Zhang2016,Bazant2004}, light~\cite{Xuan2016},
ultrasound~\cite{Ahmed2016a,McNeill2020}, or combinations thereof~\cite{Dai2016,Chen2017}. Interactions between swimmers lead to a rich variety of emergent behaviors, including self-assembly, chemotaxis, and spontaneous large-scale symmetry breaking~\cite{Palacci2013,Sen2009,Sen2019,Goldstein2007,Golestanian2007,Singh2020}.  Many of these powerful swimmers do not require toxic fuels and their compatibility with biological media enables applications in intracellular sensing and drug delivery~\cite{Wang2014,Esteban2016,Esteban2017}.

The rapid propulsion of metallic micro-rods at the nodal plane of an acoustic standing wave was first discovered in 2012~\cite{Wang2012}. 
Their electrochemical fabrication results in one end being concave and the other convex. 
With single-element rods, this shape asymmetry alone generates propulsion, but at relatively slow speeds compared to density-asymmetric bimetallic rods 
where the less dense end of the rod always leads~\cite{Ahmed2016b}. The requirement for an asymmetry in shape or density is  supported by experiments  showing locomotion  
of density-asymmetric spheres~\cite{Valdez2020} and no directional propulsion for symmetric micro-rods~\cite{Zhou2017}. 
Previous modeling proposed that acoustic propulsion of metallic rods arises from the  nonlinear inertial coupling 
between rotational and translational perturbation flows (acoustic  streaming)~\cite{Nadal2014,Lippera2019,Collis2017,Nadal2020}.
However, this model has yet to reproduce  experimental data quantitatively.

\begin{figure*}[t]
\hspace*{0cm}
\scalebox{1}{\includegraphics{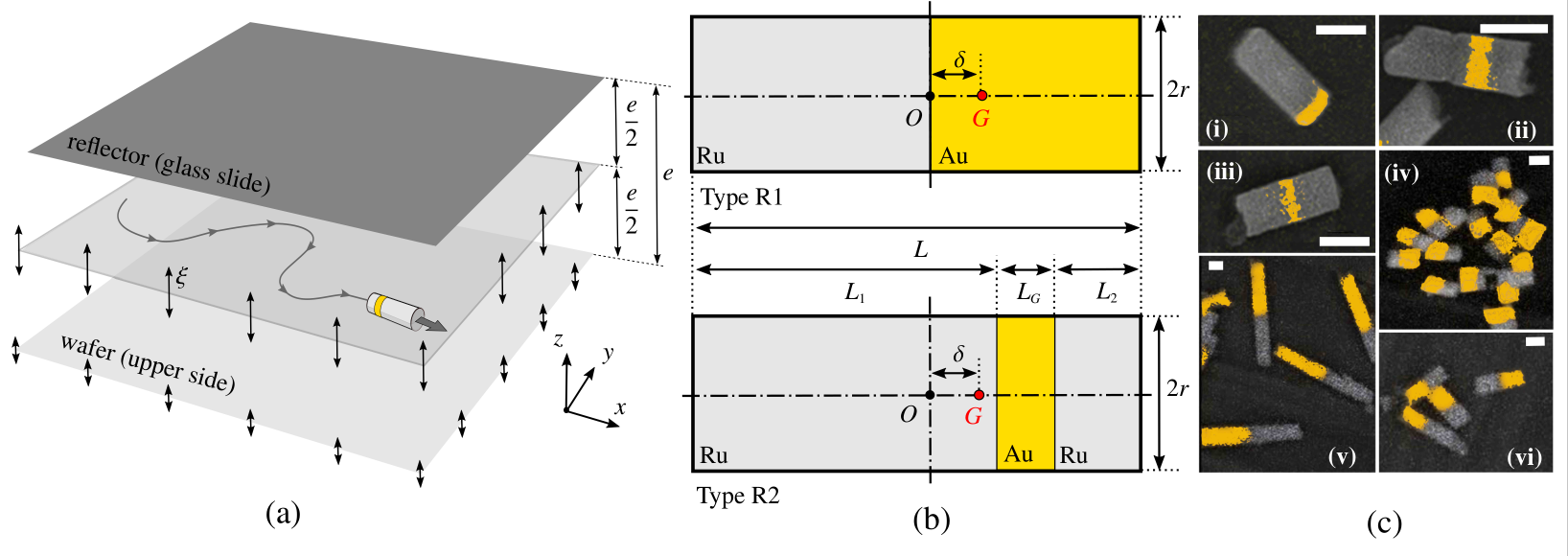}}
\caption{(a) Diagram of the acoustic resonators. The amplitude of the acoustic field displacement at the pressure nodal plane
is denoted by $\xi$ so   the corresponding transverse velocity is $\vecU_e = \xi\omega\,\exp(\icomp\omega t)\vecz$. 
(b) Geometry of the rods of types R1 and R2 . The values of $L$, $L_{\sst 1}$, $L_{\sst 2}$ and $L_{\sst G}$ and
corresponding standard deviations for type R2 are provided in  Supplementary Material~\cite{SM}. (c) SEM pictures of bimetallic
rods: (i) type R2-A; (ii) type R2-B; (iii) type R2-C; (iv) type R1, $L = 587$\,nm; (v) type R1, $L = 2458$\,nm;
(vi) type R1, $L = 990$\,nm. The contrast between the gold stripe (yellow) and the ruthenium sections (gray)
has been  enhanced. In (c)i-vi the white scale bars are 300\,$\upmu$m long.\label{fig:rods_pics}}
\end{figure*}

Here we report experiments in which the density of multi-segment bimetallic rods is varied along their length. Their forced propulsion in the nodal plane of an acoustic water-filled resonator is modeled by an unsteady viscous 
propulsion mechanism in which acoustic streaming plays no role. The amplitude of the forcing acoustic field is directly 
inferred from the drift dynamics of silica micro-spheres towards the pressure node~\cite{Barnkob2010}. 
The accurate measurement of the acoustic field displacement amplitude enables our model to
reproduce the data with no adjustable parameters, making this the first quantitative agreement with
experiments on this phenomenon.
Our   model demonstrates that imbalanced microrods powered by ultrasound are, in fact, non-reciprocal Purcell motors, analogous in their undulatory swimming cycle to {\color{black}spermatozoa}~\cite{Gadelha2020} and self-propelled viscous \emph{flapping} swimmers~\cite{Becker2003,Lauga2009,Shelley2011}.

Bimetallic micro-rods were fabricated by electrodeposition of metals into the cylindrical, 300\,nm diameter pores of anodic aluminum oxide (AAO) membranes~\cite{Wang2012, Ahmed2016b}. Two sets of rods (diameter $2r = 300$\,nm, length $L$) were made (see Fig.~\ref{fig:rods_pics}b,c): 
a first set of half-gold/half-ruthenium rods (type R1), with aspect ratio $\gamma = 2r/L$ made to vary by changing the total length $L$
(average lengths $L = 384,\,587,\,990,\,1915,\,2458$, and 3921~nm); in the second set of rods (type R2), referred to as R2-A, R2-B, and R2-C (Au stripe at the end, in the second quarter of the rod, and in the middle, respectively), the distance $\delta$ between the centroid and center of mass was varied by changing the location of the gold stripe. 
Experimentally, the lengths of the second set of rods spanned from 539 to 1284 nm, with an
average length $L = 892$\,nm and a standard deviation $\Delta L = 164$\,nm.
The dimensions of the type R1 and type R2 rods  are reported in the 
Supplementary Material~\cite{SM}.

The resonator cavity (see Fig.~\ref{fig:rods_pics}a and Supplementary Material~\cite{SM}) was filled
with deionized water 
(density $\rho$, dynamic viscosity $\eta$) seeded with microrods and a dilute suspension of 3\,$\upmu$m diameter 
polystyrene or silica spheres.
The first resonant mode frequencies, $f = \omega/(2\pi) = 2.17,\,3.17,\,3.7$ and $5.0\,\pm\,0.15$\,MHz, were obtained by sweeping the frequency of the harmonic voltage source connected to the transducer, in the region corresponding to four gap thicknesses $e = 340,\,230,\,195$ and 
$150\,\pm\,8$\,$\upmu$m. In the following, the displacement amplitude of the transverse acoustic field at the pressure nodal plane is denoted by $\xi$
so that the corresponding transverse velocity of the acoustic field is $\vecU_e = \xi\omega\,\exp(\icomp\omega t)\,\vecz$, with $\vecz$   the unit vector in the $z$-direction. Once the resonance frequency was found precisely, moving spheres were recorded rising to the pressure node when the ultrasound was turned on. The height and velocity of the tracer spheres could be inferred from the change in holographic halo diameter over time, (movie {\color{red}S1}~\cite{SM}). Rods moving in the nodal plane (movie {\color{red}S2}~\cite{SM}) were recorded for 5 seconds at 20 FPS (100 frames) and tracked using MosaicSuite with Clij GPU-acceleration \cite{clij,Moscaic} in ImageJ \cite{Fiji} to capture ensemble velocities.
 
\begin{figure*}[t]
\scalebox{1}{\includegraphics{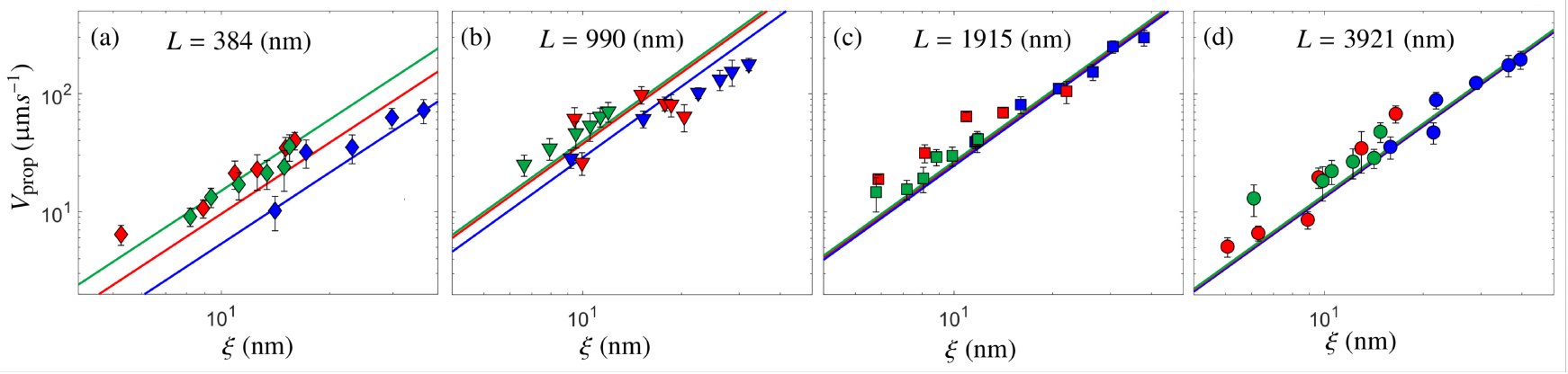}}
\vspace*{-4mm}
\caption{(a-d) Propulsion speed of type R1 rods, $\Vprop$,  as a function of the acoustic displacement amplitude at the pressure
nodal plane, $\xi$ (nm),  for four different rod lengths. The experimental data are plotted in solid symbols. {\color{black} The viscous
model depicted in the text (Eq.\,\ref{eq:sol}), for which $\Vprop\sim \xi^2$, is shown in solid lines}. Blue, red and green symbols/lines correspond to the frequencies $f = 2.17,\,3.17$ and $5.0$ MHz respectively. \label{fig:Vprop_xi}}
\end{figure*}
\begin{figure}[b]
\vspace*{-0.5cm}
\hspace*{0mm}
\scalebox{1}{\includegraphics{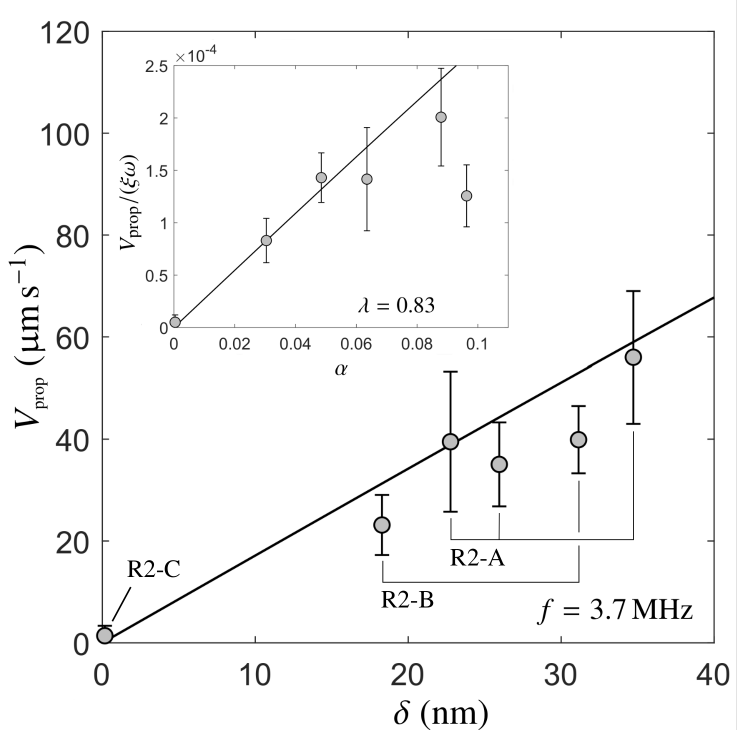}}
\caption{Propulsion speed of type R2 rods,  $\Vprop$ ($\upmu$m\,s$^{-1}$), as a function of the asymmetry distance, $\delta$ (nm), for $f = 3.7$\,MHz. Due to fluctuations in the manufacturing process
the length of type R2 rods lies in the range 539 - 1284\,nm. The
theory (viscous model) is plotted with a solid line for the mean value of the length $L = 892$ nm. 
Inset: dimensionless form of the plot.\label{fig:Vprop_alpha}}
\end{figure}%

As in past work~\cite{Wang2012,Ahmed2016b},  the lighter end of the rods (Ru side)
was always leading during locomotion (Fig.~1 in Supplementary Material \cite{SM}).
The propulsion speed  of the R1 rods, $\Vprop$,  is plotted in Fig.~\ref{fig:Vprop_xi} as a function of the displacement
amplitude, $\xi$, for {\color{black} four rod lengths} and three forcing frequencies. The whole set of data can be fitted to a power law  $\Vprop \sim \xi^{1.75}$.  
The propulsion speed of R2 rods observed for a forcing frequency $f = 3.7$\,MHz is shown  in Fig.~\ref{fig:Vprop_alpha} as a function
of the  distance  $\delta$ between the geometric center and the center of mass (see Fig.~\ref{fig:rods_pics}b). 
As expected, $\Vprop$ is an increasing function of $\delta$,
which is the controlling symmetry-breaking parameter for type R2 rods. 
In the following, we show that the propulsion speed observed for both R1 and R2 rods can be
reproduced quantitatively by a  viscous mechanism in which only the inertia of the solid (and not the fluid) is accounted for.

Using a general dimensionless analysis, we see that the normalized acoustic propulsion speed of a rigid rod, $\Vprop/(\xi\omega)$, 
depends on six dimensionless parameters \cite{Nadal2020}: the dimensionless asymmetry $\alpha = 2\delta/L$; 
the fluid-to-solid density ratio $\beta = \rho/\bar{\rho}$ ($\bar{\rho}$ being the mean density of the rod); the aspect ratio $\gamma = 2r/L$;
the frequency Stokes parameter (which can be interpreted as the inverse of the
dimensionless viscous diffusion length) $\lambda = (\rho\,r^2\omega/\eta)^{1/2}$; the dimensionless amplitude of the acoustic field
at the pressure nodal plane $\veps = \xi/r$; and the dimensionless inertia $\tilde{I} = I/I_h$ ($I$ and $I_h$ being the inertiae of
the original rod and that of the iso-volume homogeneous rod about their respective centers
of mass). 

\begin{figure}[t]
\hspace*{0cm}
\scalebox{1}{\includegraphics{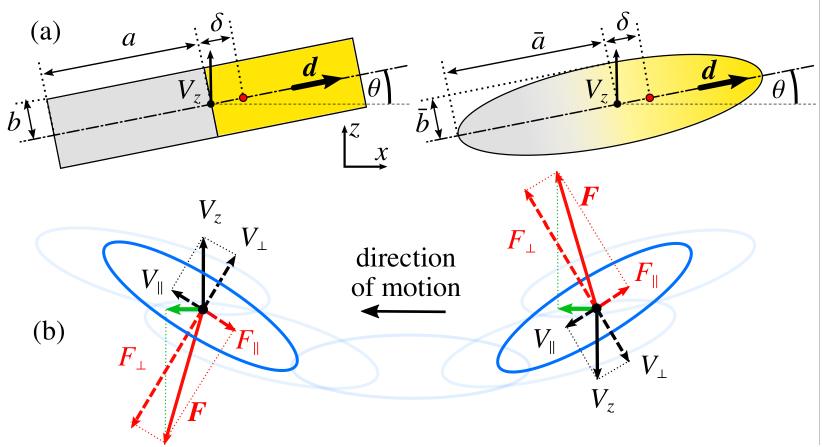}}
\vspace*{-2mm}
\caption{(a) Equivalent ellipsoid model of the cylindrical rods used to derive an expression of the
propulsion speed $\Vprop$. (b) Physical principle of the rectified viscous mechanism  
based on the combined effects of  shape anisotropy and the   flapping motion~\cite{Was2014} . Regardless of the direction of
the vertical velocity within a period of oscillation, the shape anisotropy yields a rectified force oriented towards the less dense end
of the solid (green arrow).
\label{fig:model}}
\end{figure}

Past studies argued that the acoustic propulsion of metallic rods resulted from the nonlinear inertial coupling in the fluid between rotation and translation (phenomenon of so-called acoustic streaming). However, this physical mechanism has not been tested quantitatively, due to lack of knowledge of the local acoustic displacement amplitude. We used direct measurements of $\xi$ and introduced reasonable values of the
previously proposed dimensionless parameters in the available streaming-based predictive
theories~\cite{Nadal2014,Lippera2019,Collis2017,Nadal2020}. This led to quantitative predictions of the propulsion speed that underestimated
the experimental values by more than one order of magnitude (see Supplementary Material~\cite{SM}). 
The question therefore remains to determine the main physical mechanism governing acoustic propulsion. We argue here that the propulsion mainly results from the {non-reciprocal (rectified) flapping} motion
of the rods, a viscous mechanism  originally proposed in Ref.~\cite{Was2014} for ellipsoids and illustrated in
Fig.~\ref{fig:model}b (see theoretical section below). In this mechanism, a propulsive force is created if a time-varying 
displacement of an elongated body (flapping) is combined with a time-varying change of conformation (angle) in a non-time-reversible fashion. To adapt these earlier results~\cite{Was2014}, we  consider prolate spheroids of the same 
mass and aspect ratio (semi-minor and major axis $\bar{b} = [3\gamma m/(4\pi\bar{\rho})]^{1/3}$
and $\bar{a} = \bar{b}/\gamma$, see Fig.~\ref{fig:model}a) whose flapping motion is forced by the transverse acoustic field. 
Note that this leads to a re-definition of the frequency parameter $\lambda = \bar{b}^2\omega/\nu$ and dimensionless amplitude
$\veps = \xi/\bar{b}$, which are now based on the semi-minor axis $\bar{b}$, while the remaining  dimensionless parameters are
those of the initial rod. The averaged propulsion speed $\Vprop $ derived in Ref.~\cite{Was2014} (where $\langle\cdot\rangle$
denotes the average over a period of oscillation) is 
\begin{equation}
\Vprop = \frac{\Lperp - \Lpar}{\Lpar}\langle V_z\,\theta\rangle,
\label{eq:viscous_prop_vel}
\end{equation}
from which one can infer that, in order to obtain a non-zero  mean speed: (i) the local transverse 
(along the minor axis) and lengthwise (along the major-axis) Stokes 
drag coefficients $-6\pi\bar{a}\Lperp$ and $-6\pi\bar{a}\Lpar$ must be different, and (ii) the vertical 
velocity $V_z$ of the centroid (relative to the fluid) and the tilt angle $\theta$ of the major-axis must
not be $\pi/2$ out-of-phase from one another (otherwise the average would be zero). 

In order to compute the propulsion speed, $\Vprop$,  we calculate the time-periodic quantities
$V_z$ and $\theta$  from Newton's laws by taking into account the inertia of the solid (only)~\cite{Nadal2020}.
We consider the incompressible limit in which the transverse dimensions of the rod are  small compared to the acoustic wavelength.
In this case, the forcing oscillatory flow can be approximated as locally uniform at the particle scale. In addition, $\xi$
is assumed to be smaller than the minor axis of the rod, so $\veps\ll 1$.  
We write the \emph{imbalance} vector of the ellipsoid that results from its density asymmetry as $\vecdel = \mathbf{OG} = \delta\,\vecd$
where $\vecd$ is the unit vector aligned with the major axis of the ellipse (Fig.~\ref{fig:model}). The conservation
of momentum can be linearized in the small tilt angle limit ($\theta \ll 1$) as
$\bar{\rho}\,\mathcal{V}\dvecV = m\,\vecdel\times\ddth
+\,\vecF + (\rho-\bar{\rho})\mathcal{V}\,\dot{\vecU}_e$ where $\ddth = \ddot{\theta}\,\vecy$, $\vecV = V_z\,\vecz$
is the oscillating velocity of the solid relative to the surrounding fluid, $\mathcal{V}$ is the volume of the particle,
$\vecF$ is the corresponding viscous hydrodynamic force, and the last term is the effective buoyancy force.
Similarly, the conservation of  angular momentum can be written as $I\,\ddth =  \vecL - \vecdel \times (\vecF + \rho\mathcal{V}\dot{\vecU}_e)$,
where $\vecL = -8\pi\bar{a}^3\Lr\,\dth$ is the hydrodynamic torque about the $(O,y)$-axis due to rotation.
Using $\xi \omega$ and $\omega^{-1}$ as reference velocity and time scale, and denoting by $v_z$ the
dimensionless velocity $V_z/(\xi\omega)$ yields the  
non-dimensional forms of {\color{black} momenta} conservation equations 
\begin{gather}
\dot{v}_z = -\frac{A}{\veps}\,\ddot{\theta} - B\,v_z + E\,\expo^{\icomps t}\label{eq:sys1},\\
\ddot{\theta} = - C\,\dot{\theta} + \veps D\,v_z - \veps F\,\expo^{it}\label{eq:sys2},
\end{gather}
 where  $A = (\alpha/\gamma)$, $B = [(9\beta\Lperp/(2\lambda^2)]$, $C = [15\Lr/(\lambda^2\tilde{I}\chi)]$, $D = [9\beta\Lperp/(2\lambda^2)]$, $E = \icomp(\beta-1)$ and $F = \icomp[5\alpha\beta\gamma/(2\tilde{I}\chi)]$ do not depend on $\veps$, since $\chi  = (1+\gamma^2)/2$, $\Lpar$, $\Lperp$ and $\Lr$
only depend on the aspect ratio of the rod.
From Eqs.~(\ref{eq:sys1}-\ref{eq:sys2}), it can be inferred that $v_z$ and $\theta$ are $O(1)$ and $O(\veps)$
respectively. So, writing these two quantities as $v_z = \hat{v}_{\sst 0}\expo^{\icomps\omega t}$ and 
$\theta = \veps\,\hat{\theta}_{\sst 0}\expo^{\icomps\omega t}$, solving  the system  in Eqs.~(\ref{eq:sys1}-\ref{eq:sys2}) for $\hat{v}_{\sst 0}$ and $\hat{\theta}_{\sst 0}$, and  evaluating Eq.~(\ref{eq:viscous_prop_vel}) leads to our theoretical prediction for the propulsion speed
\begin{equation}
\Vprop = \frac{1}{2}\xi\omega\,\veps\,\Re\left[\hat{\theta}_{\sst 0}^\dag\hat{v}_{\sst 0}\right].\label{eq:sol} 
\end{equation}
The  speed is thus of the form $\Vprop = \xi\omega\,\veps\,\mathcal{F}(\alpha,\beta,\gamma,\lambda,\tilde{I})$ - 
where the dimensionless factor $\mathcal{F} = (1/2)\Re[\hat{\theta}_{\sst 0}^\dag\hat{v}_{\sst 0}]$ does not depend on $\veps$, and is  
therefore proportional to the square of the forcing amplitude $\xi$. {\color{black} This quadratic dependence is close to the experimental scaling $\Vprop\sim \xi^{1.75}$ obtained from a global least square fitting of the data 
(in Fig.~\ref{fig:Vprop_xi}, theory is plotted in solid lines)}.    

\begin{figure}[t]
\scalebox{1.05}{\includegraphics{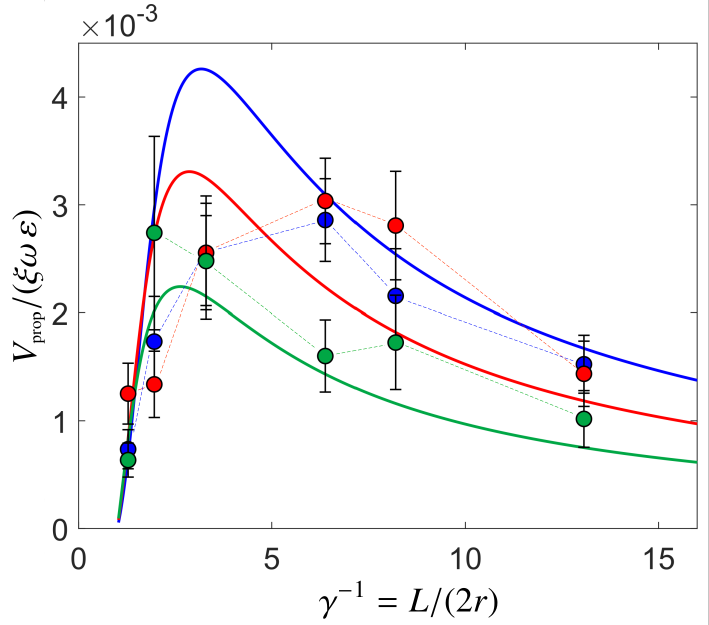}}
\vspace{-2mm}
\caption{Dimensionless propulsion velocity $\Vprop/(\xi\omega)$ for type R1 rods rescaled by the dimensionless field amplitude $\veps$. The experimental data
are plotted in solid symbols; the theory is plotted in solid lines. The colour code is the same as in Fig.~\ref{fig:Vprop_xi}.\label{fig:rescaled_Vprop}} 
\end{figure}

The theoretical predictions are also shown in Fig.~\ref{fig:Vprop_alpha} in both dimensional and dimensionless forms. For small
values of the asymmetry parameter $\alpha$, one expects a linear dependence of the propulsion velocity with 
$\Vprop = \xi\omega\,\veps\,\alpha\,[\partial_\alpha\mathcal{F}]_{\sst \alpha = 0}$, as obtained. In both cases,
the propulsion speed is quantitatively captured by the model, with no  fitting parameters.
The agreement includes both the direction of swimming (light end of the rod leading) and its magnitude.
In Fig.~\ref{fig:rescaled_Vprop}, we next plot the dimensionless function $\mathcal{F}$   as a function of the inverse of the aspect ratio (dimensionless length)
from the data in Fig.~\ref{fig:Vprop_xi}, by averaging the quantity $\Vprop/(\xi\omega\,\veps)$ over $\veps$ for each
pair of parameters $(\lambda,\gamma)$, i.e., for each pair of dimensioned parameters $(f,L)$. Note that,
for type R1 rods, $\mathcal{F}$ is a function of $\gamma$ and $\lambda$ only, since the values of the constants 
$\alpha \simeq 0.1$, $\beta \simeq 0.06$ and $\tilde{I} \simeq 1$ do not depend on the aspect ratio.
{\color{black}Small discrepancies exist,  e.g.~the optimal length, which can be estimated from the 
measurements for $\lambda = 0.55$ and $\lambda = 0.67$ ($f = 2.17$ and $3.17$\,MHz), 
is underestimated by the model (at least by a facor 2) and so is the corresponding 
value of function $\mathcal{F}$, but despite these, the model captures the major trends and orders of magnitude
observed in the experimental data}. In particular, the drop in 
propulsion velocity observed for $\gamma\rightarrow 1$ (due to the loss of anisotropy) agrees with the  model,
as is the decrease observed for $\gamma^{-1} \gg 1$, where axial drag becomes prominent.

In summary, we have presented here  evidence  that the acoustic propulsion of
bimetallic cylindrical rods is driven by a viscous mechanism in which the fluid inertia plays no role.  
Mathematically, ignoring  the inertia of the fluid  is  a  valid assumption provided $\lambda^2 \ll 1$; in our experiments the values of $\lambda^2$ range from
0.3 to 0.7, and therefore our model remains quantitively accurate even beyond the strict asymptotic limit in which it should be valid. 
Note that our model  does not rule out contributions from other streaming-based mechanisms, which remain
relevant in the case of pure spheres with strong imbalance \cite{Valdez2020}. The agreement between the model and our data with no fitting parameters strongly suggests that viscous flapping is the main mechanism governing the physics of acoustic propulsion of density-asymmetric metallic rods. Our work represents therefore a paradigm shift
in our understanding of acoustic propulsion. With more detailed modeling and  complex nanorod designs, a full description of this phenomenon  should be within reach. These insights could inform the rational design of many different types
of future nanoscopic swimmers and active matter powered by ultrasound and represent an important  step towards a more wholistic understanding of
ultrasound-matter interactions and their nanoscale fluid dynamics, with potential impact in biology, medicine,
sensing, and fluid physics. 

This project received funding from the European Research Council (ERC) under the European Union's Horizon 2020 research and innovation programme (grant agreement 682754 to EL). N.S. acknowledges Research Experiences for Undergraduates support under National Science Foundation grant DMR-1952877. This work was carried out in part at the Singh Center for Nanotechnology, which is supported by the NSF National Nanotechnology Coordinated Infrastructure Program under grant NNCI-2025608.

\bibliography{bib_rods}

\appendix
\cleardoublepage
\pagebreak
\newpage
\section{SUPPLEMENTARY MATERIAL}

\subsection{Measurement of the acoustic field amplitude}
\begin{table*}[h!]
\begin{tabular}{C{0.06\textwidth} C{0.15\textwidth} C{0.15\textwidth} C{0.15\textwidth} C{0.15\textwidth} C{0.15\textwidth} C{0.15\textwidth}}
\toprule
& {\bf\boldmath R2-A (batch \#1)} & {\bf\boldmath R2-A (batch \#2)} & {\bf\boldmath R2-A (batch \#3)} & {\bf\boldmath R2-B (batch \#1)} & {\bf\boldmath R2-B (batch \#2)} 
& {\bf\boldmath$C$}\\\hline
{\bf\boldmath$\xi$\,(nm)} & 10.3 & 10.3 & 12.9 & 14.4 & 12.1 & 10.5 \\\bottomrule
\end{tabular}
\caption{Fluid displacement amplitude $\xi$ at the pressure nodal plane $(z = 0)$ obtained for type R2 rods ($f = 3.7$\,MHz).
The corresponding average value is $\langle\xi\rangle = 12.1$\,nm.
\label{tab:xi}}
\end{table*}
The amplitude $\xi$ of the acoustic field - i.e. the displacement amplitude of the fluid particles along the 
$z$-direction in the vicinity of the nodal pressure plane
($z = 0$) - is an important parameter of the problem since it determines the intensity of the transverse hydrodynamic
forcing experienced by the micro-rods. A direct measurement of $\xi$, for instance by means of Particle Image
Velocimetry (PIV), is not relevant since at such frequencies ($f = 3.7$\,MHz) and such small displacements ($\xi$ is
of the order of few tens of nanometers at most), PIV tracers are no longer passive. As an alternative, we exploited
the drift of polystyrene particles (radius $a = 1.5$\,$\upmu$m, density $\rho_p = 1.05\times 10^3$\,kg\,m$^{-3}$, speed of sound
in polystyrene $c_p \simeq 1700$\,m\,s$^{-1}$) towards the nodal pressure plane under the effect of the acoustic force, as suggested
in Ref.\,\cite{Barnkob2010}.
At the beginning of each experiment - i.e.~for each type of rod - the measurement of the drift of the polystyrene particles
was performed by analyzing the change of the de-focused optical pattern upon actuation of the acoustic field
in a region directly located below the plane in which the self-propelled bimetallic rods
were observed. The amplitude of the acoustic velocity field at the pressure node was then inferred as follows. 
In the resonator (gap thickness $e$) the acoustic force experienced by a polystyrene particle is given by
\begin{equation}
\vecF = \frac{\pi}{3}\Phi\,F^\star\,\sin(2kz')\,\vecz,
\label{eq:acoustic_force}
\end{equation}
where $z' = z + e/2$ is the height of the particle relative to the bottom wall and $\vecz$ is the unit vector along the $z$-direction.
In the expression above, $F^\star = \rho(\xi c)^2(k a)^3$ is the acoustic force scale, and the compressibility
factor $\Phi$, which is a function of the density ratio $\beta_p = \rho/\rho_p$ and the speed of sound ratio
$\chi_p = c/c_p$, have the following form 
\begin{equation}
\Phi = \frac{5-2\beta_p}{2+\beta_p} - \beta_p \chi_p^2.
\label{eq:comp_coef}
\end{equation}
Taking $\rho = 9.98 \times 10^3$\,kg\,m$^{-3}$ for the density of water yields $\Phi = 0.328$.

Balancing the acoustic force defined above and the viscous drag $\vecF_{\sst \!\!D} = -6\pi\eta\,a\,\dot{z}'\,\vecz$, 
and ignoring the inertia and effective weight
of the polystyrene particles (since $\rho \simeq \rho_p$) leads to the following form of momentum 
conservation along the $z$-direction:
\begin{equation}
\dot{z}' = Y \sin(2 k z')\;\;\mbox{with}\;Y = \frac{\Phi F^\star}{18\,\eta\,a}.\\
\label{eq:acoustic_force_simpl}
\end{equation}
\vspace{1mm}

The solution to the first order differential equation above is of the form $z'(t) = k^{-1}\,\cot^{-1}[\exp[-2k(K+Yt)]]$,
where $K$ is a constant of integration. Both parameters $K$ and $Y$ were adjusted to best fit the experimental
time profile $z'(t)$ measured for each type of rod. As an example, the values found in the case
of type R2 rods, i.e.~for a frequency $f = 3.7$\,MHz, were consistently found in the range [10.3\,nm\,$-$\,14.4\,nm] (see Table\,\ref{tab:xi}), 
leading to a mean value of 12.1\,nm. The values of $\xi$ found in each experimental run were
used to assess the quantitative validity of our proposed viscous model and of the
inertia-based models available in the literature \cite{Nadal2020,Collis2017}.

\subsection{Fabrication method and dimensions of type R1 and type R2 rods}

The bimetallic cylindrical rods were grown via templated electrodeposition of gold (density 
$\rho_{g} = 19.32\,\times\,10^3$\,kg\,m$^{-3}$) and ruthenium (density 
$\rho_{r} = 12.45\,\times\,10^3$\,kg\,m$^{-3}$) into the pores of an anodic aluminum oxide (AAO) membrane of 300\,nm pore diameter 
- with 400\,nm of Ag thermally evaporated onto the branched side of the membrane - by controlling the plating time and current. 
Silver, gold, and ruthenium were deposited from commercial plating solutions (Technic, USA) by making 
electric contact between the silver side of the membrane and a strip of copper
tape (serving as the cathode) \cite{Ahmed2016b}.  A platinum wire submerged in the plating solution served as the
anode. Using a potentiostat/galvanostat (NuVant Systems EZstat) to control the current, the
branched part of the membrane was filled with silver by depositing at $- 6$\,mA ($- 1.22$\,mA\,cm$^{-2}$)
for 30 minutes, and gold and ruthenium were deposited at $- 0.61$\,mA\,cm$^{-2}$ and $- 2.04$\,mA\,cm$^{-2}$ respectively, giving deposition rates of
100 nm/min and 40 nm/min, respectively. By this method, nanorods of arbitrary segment length
could be fabricated in large numbers. The silver backing was then dissolved in 10\,M nitric acid for 15
minutes, followed by a thorough rinse with de-ionized (DI) water and submersion of the membrane in 5\,M NaOH
for 30 minutes to dissolve the alumina template. The rods were rinsed with DI water and centrifuged three times to
ensure removal of contaminants before suspending them in 20\,mL of DI water for acoustic actuation.

Type R1 rods consisted of two sections (Au and Ru) of equal lengths $L/2$, as depicted in Fig.\,1b of the main article. 
The general features of the type R2 micro-rods (see again Fig.\,1b in the main article) were as follows: a denser gold stripe of length
$L_{\sst G}$ was inserted between two ruthenium sections of respective lengths $L_{\sst 1}$ and $L_{\sst 2}$. 
The initial objective was to keep the total length $L = L_{\sst 1} + L_{\sst 2} + L_{\sst G}$ 
and that of the gold stripe (i.e. $L_{\sst G}$) constant, and only modify the position of the stripe along 
the rod axis. Doing so, we intended to change the ``intensity" of the left-right symmetry 
breaking, which can be quantified by the normalized quantity $\alpha = 2\delta/L$, where $\delta$ is 
the distance between the centroid $O$ and the center of mass $G$ of the rod. We actually fabricated
three different kinds of type R2 rods (labeled as A, B and C) with same diameter $2r = 300 \pm 25$\,nm 
and different imbalance coefficients $\alpha$; however, due to inevitable fluctuations in the fabrication process, 
the length of the gold stripe and the total length were not exactly the same from one type to another. 

The geometrical characteristics of each type of rods are reported in Tables \,\ref{tab:dimensional_geom_R1} and \,\ref{tab:dimensional_geom_R2}. 
Note that for type R2-A rods the gold stripe was located at one end, so that 
in this case $L_{\sst 2} = 0$\,nm. The standard deviation for each length reported in the tables, as well as for the rod diameter,
is around 10\%.

\begin{table*}[h!]
\begin{tabular}{C{0.18\textwidth} C{0.05\textwidth} C{0.05\textwidth} C{0.05\textwidth} C{0.05\textwidth} C{0.05\textwidth}
C{0.05\textwidth} C{0.05\textwidth} C{0.05\textwidth} C{0.05\textwidth} C{0.05\textwidth} C{0.05\textwidth} C{0.05\textwidth}}
\toprule
\bf\boldmath Total length $L$\,(nm) & \multicolumn{2}{c}{\bf 384} & \multicolumn{2}{c}{\bf 587} &	\multicolumn{2}{c}{\bf 990}	& 
\multicolumn{2}{c}{\bf 1915} & \multicolumn{2}{c}{\bf 2458} & \multicolumn{2}{c}{\bf 3921}\\
 & \bf Ru & \bf Au & \bf Ru & \bf Au & \bf Ru & \bf Au & \bf Ru & \bf Au & \bf Ru & \bf Au & \bf Ru & \bf Au \\
\cmidrule(lr){2-3} \cmidrule(lr){4-5} \cmidrule(lr){6-7} \cmidrule(lr){8-9} \cmidrule(lr){10-11} \cmidrule(lr){12-13}
\bf\boldmath Au/Ru lengths (nm) & 178 & 206 & 283 & 305 & 505 & 485 & 900 & 1015 & 1211 & 1247  & 1769 & 2152 \\
\bottomrule
\end{tabular}
\caption{Geometric and physical characteristics of type R1 rods, as depicted in the main article.  
\label{tab:dimensional_geom_R1}}
\end{table*}

\begin{table*}[h!]
\begin{tabular}{C{0.16\textwidth} C{0.08\textwidth} C{0.08\textwidth} C{0.08\textwidth} C{0.08\textwidth} C{0.08\textwidth} C{0.17\textwidth} C{0.13\textwidth}}
\toprule
 {\bf Type}  & {\bf\boldmath$L_{\sst 1}$\,(nm)} & {\bf\boldmath$L_{\sst G}$\,(nm)} &
 {\bf\boldmath$L_{\sst 2}$\,(nm)} & {\bf\boldmath$L$\,(nm)} & {\bf\boldmath$\delta$\,(nm)} & {\bf\boldmath$I$\,(kg\,m$^2$) $\times 10^{27}$} & {\bf\boldmath$\bar{\rho}$\,(kg m$^{-3}$) $\times 10^{4}$}\\[1mm]\hline
 {\bf R2-A (batch \#1)}  & 614 & 102 &  0  & 717 & 23 & 0.0313 & 1.34\\
 {\bf R2-A (batch \#2)}  & 609 & 181 &  0  & 790 & 35 & 0.0435 & 1.40\\
 {\bf R2-A (batch \#3)}  & 393 & 146 &  0  & 539 & 26 & 0.0139 & 1.43\\
 {\bf R2-B (batch \#1)}  & 688 & 220 & 295 & 1203 & 18 & 0.1311 & 1.36\\
 {\bf R2-B (batch \#2)}  & 571 & 493 & 220 & 1284 & 31 & 0.1653 & 1.50\\
 {\bf R2-C}  & 347 & 130 & 344  & 820  	& 0.1 & 0.0406 & 1.35\\\bottomrule
 \end{tabular}
\caption{Geometric characteristics of type R2 rods, as depicted in the main article. Note that for type R1 rods, the dimensionless imbalance 
$\alpha$ is consistently close to 0.1.\label{tab:dimensional_geom_R2}}
\end{table*}

\subsection{Design and fabrication of the acoustic resonators}

The acoustic actuator consisted of a piezo transducer (STEMiNC, model SMD12T06R412WL, USA)
epoxied to the unpolished side of a piece of a silicon wafer.
Directly above the transducer, a number of  Kapton annular spacers of different thicknesses 
were sandwiched between the opposite (polished) side of the wafer and a glass slide (used as an acoustic reflector),
leading to different gap thickness $e = 340,\,230,\,195$ and $150\,\pm\,8$\,$\upmu$m.
The resonator cavity was filled with $\sim$\,6\,$\upmu$L of water (density $\rho = 998$\,kg\,m$^{-3}$,
viscosity $\eta = 10^{-3}$\,Pa\,s) that contained a mixture of bimetallic micro-rods, and 3\,$\upmu$m diameter
polystyrene spheres - the purpose of the latter being to enable a measurement of the acoustic field amplitude.
After sweeping the frequency of the harmonic voltage signal
delivered by a waveform generator (Siglent SDG2122X) hooked to the
leads of the transducer, the first resonant mode conditions were found around $f = \omega/(2\pi) = 2.17,\,3.17,\,3.7$ and $5.0\,\pm\,0.15$\,MHz
respectively, for the gap thicknesses listed above.
In practice, this frequency corresponds to the configuration where the gap thickness equals
half a wavelength $\lambda = c/f$ (where $c = 1480$\,m\,s$^{-1}$ is the speed of sound in water
at ambient temperature). At the resonant frequency, the micro-rods and the spheres gather into the single 
zero pressure plane located halfway between the silicon wafer and the glass slide.

\subsection{Direction of locomotion for type R1 rods}
\begin{figure}[t]
\includegraphics[width=0.49\textwidth]{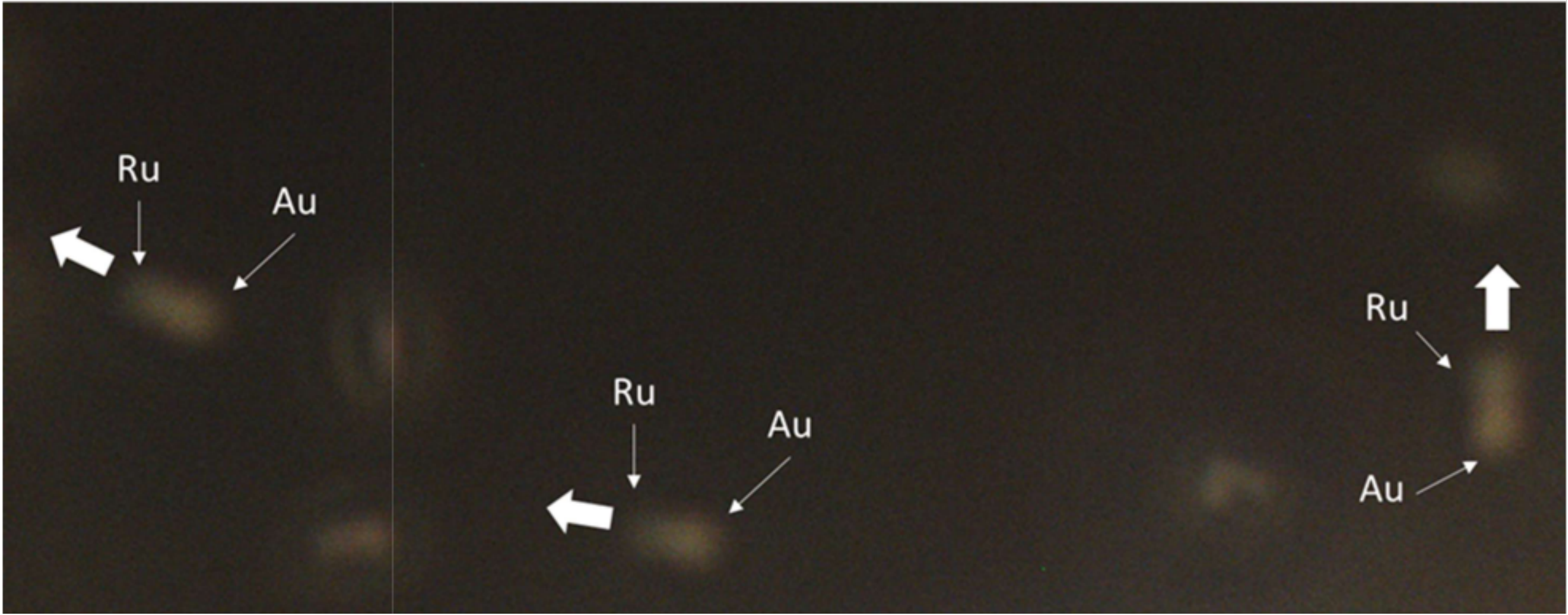}
\caption{Direction of propulsion of type R1 rods ($L=990$\,nm). Despite the poor optical contrast, it is possible to
distinguish the Au end from the Ru end. Note that this is not possible in the case of type R2 rods.}
\label{fig:prop_dir}
\end{figure}
Despite the poor optical contrast between Ru and Au ends, we were able to determine with certainty in
which direction type R1 micro-rods were swimming, as shown in Fig.\,\ref{fig:prop_dir} As in the cases reported by
Wang {\it et al.} \cite{Wang2012} and Ahmed {\it et al.}, the lighter end is leading - i.e. the end that is the farthest from the
center of mass \cite{Ahmed2016b}.\\[2mm]

\subsection{Order of magnitude produced by streaming based models}
\begin{figure}[t]
\scalebox{0.66}{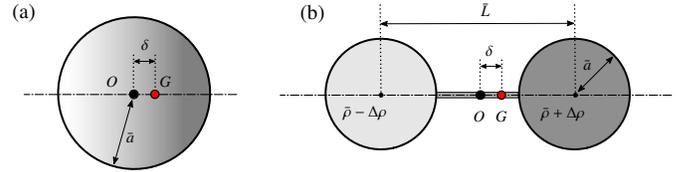}
\caption{Sketches of the equivalent systems considered by (a) Collis {\it et al.} \cite{Collis2017} and
(b) Nadal \& Michelin \cite{Nadal2020} to model inertia-based acoustic propulsion. Darker shades of gray
correspond to denser regions. In the dumbbell model proposed by Collis, each sphere is homogeneous in density
and $L\gg \bar{a}$.
\label{fig:model_equiv}}
\end{figure}
In previous works \cite{Valdez2020,Ahmed2016b,Wang2012}, measurements of the
acoustic field amplitude $\xi$ were unavailable.  This has so far prevented a quantitative comparison between the experimental results
reported and the extant streaming-based inertial theories. By filling this gap we are now in position to quantitatively discuss
these models based on more realistic values of the forcing - i.e. the fluid $z$-displacement amplitude at the nodal plane.\\

\paragraph{Equivalent mass bottom-heavy sphere} - We first consider the model proposed by Nadal \& Michelin \cite{Nadal2020}.
An equivalent mass sphere - i.e. a sphere with the same mean density $\bar{\rho}$ and same volume $\mathcal{V}$ - is considered for each different
type of rod (see Fig.\,\ref{fig:model_equiv}a). The radius $\bar{a} = [(3/4)\mathcal{V}/\pi]^{1/3}$ of the equivalent sphere is used to re-compute an equivalent 
dimensionless frequency $\bar{\lambda} = \bar{a}^2\omega/\nu$ and an equivalent dimensionless 
forcing parameter $\bar{\varepsilon} = \xi/\bar{a}$. For each equivalent sphere, the imbalance parameter
$\alpha$, the dimensionless inertia $\tilde{I}$ and the density ratio $\beta$ are those of the initial rod.
As an example, introducing these parameters together with the measured amplitude of the acoustic field amplitude $\xi$ in the model presented in Ref.\,\cite{Nadal2020}
for type R2-A\,(batch \#1), R2-A\,(batch \#2) and R2-A\,(batch \#3) rods yields to $\Vprop = 2.1$\,$\upmu$m\,s$^{-1}$,  $\Vprop = 2.8$\,$\upmu$m\,s$^{-1}$ and $\Vprop = 3.05$\,$\upmu$m\,s$^{-1}$ respectively for the value of the propulsion speed, which systematically underestimates the measurements by at least one order of magnitude.
A similar discrepancy also exists for R2-B rods and large aspect ratio type R1 rods.\\

\paragraph{Equivalent mass asymmetric dumbbell} - In the same way, an equivalent mass dumbbell similar to those
considered by Collis {\it et al.} is built from each initial rod. The obtained equivalent (common) radius of the
sphere $\bar{a}$ is used to compute an equivalent length $\bar{L} = 2\,[I/m -(2/5)\,\bar{a}^2]^{1/2}$ (distance between sphere centers)
leading to the same inertia as that of the initial rod. The equivalent length $\bar{L}$ is then used to calculate the density difference
$\Delta\rho$ leading to a value of the imbalance parameter $\alpha$ equivalent to that of the initial rod.
The equivalent densities of the left and right spheres in Collis' model are then given by 
$\rho_{\sst 1} = \bar{\rho} + \Delta\rho$ and $\rho_{\sst 2} = \bar{\rho} - \Delta\rho$ with $\Delta\rho \sim \alpha \bar{\rho}$.
Collis' density parameter $(\bar{\rho} - \Delta\rho -\rho)/(\bar{\rho} + \Delta\rho - \rho)$ and an equivalent
dimensionless frequency $\bar{\lambda} = \bar{a}^2\omega/\nu$ are calculated from the equivalent radius and density difference.
Finally, the normalized dimensionless quantity $\hat{U}_{prop} = (\bar{L}/\bar{a})\, U_{prop}$ is read from Fig.2 in Ref.\,\cite{Collis2017}.
Considering the values for $\bar{\lambda}$ and Collis' density parameters, which systematically lie in the range [0.8-0.9]
for type R2-A and R2-B rods, a generic value $\hat{U}_{prop} = 5\times 10^{-4}$ is considered for all cases as an upper boundary. Introducing
the measured value of the acoustic field amplitude displacement $\xi$ in the expression of the dimensional propulsion 
velocity ($\Vprop = \xi^2\omega/\bar{a}\,U_{prop}$) leads to $\Vprop = 2.95$\,$\upmu$m\,s$^{-1}$, 
$\Vprop = 2.8$\,$\upmu$m\,s$^{-1}$ and $\Vprop = 4.23$\,$\upmu$m\,s$^{-1}$ respectively for type R2-A\,(batch \#1), R2-A\,(batch \#2) and R2-A\,(batch \#3) rods.
As expected, since the same physical mechanism is involved, the orders of magnitude are similar to those produced by the model in Ref.\,\cite{Nadal2020}.\\[1mm]

In addition to the underestimation of the propulsion speed for large aspect ratio rods, acoustic streaming models do not capture the 
observed prominent drop of the propulsion speed for $O(1)$ aspect ratio, which can be straightforwardly verified for the model presented
in Ref.\,\cite{Nadal2020}. Despite the substantial difference of shape between the models proposed in Refs.\,\cite{Collis2017,Nadal2020} 
(bottom-heavy sphere and density assymetric dumbbell) and the real rods, the information drawn from a comparison with experiments
strongly suggests that streaming is not the dominant mechanism at work in the case of bimetallic rods. This is the reason why
an alternative purely viscous mechanism is also proposed, which fits the experimental results without any adjustable parameters. Note that
the proposed mechanism cannot explain the acoustic propulsion observed for strongly imbalanced spheres reported in Ref.\,\cite{Valdez2020}.
In that case, the streaming-based mechanisms remain relevant and are the best candidates to interpret the observed acoustic propulsion.

\vfill
\end{document}